\documentclass[12pt]{iopart}
\usepackage{graphics}
\usepackage{graphicx}
\usepackage{epsfig}
\usepackage{amssymb}
\usepackage{amsfonts}
\usepackage{color}

\providecommand{\U}[1]{\protect\rule{.1in}{.1in}}
\begin{document}

\title{Mechanism for enhancement of superconductivity in multi-band systems with odd parity hybridization}

\author{Mucio A. Continentino$^1$, Igor T. Padilha$^2$ and  Heron Caldas$^3$} 

\address{$^1$ Centro Brasileiro de Pesquisas Fisicas, Rua Dr. Xavier Sigaud, 150, Urca \\
22290-180, Rio de Janeiro, RJ, Brazil}

\address{$^2$ Universidade Federal do Amazonas \\
Campus Capital, 69077-070, Manaus, AM, Brazil}

\address{$^3$ Departamento de Ci\^{e}ncias Naturais, Universidade Federal de
  S\~ao Jo\~ao Del Rei, \\ 36301-000, S\~ao Jo\~ao Del Rei, MG, Brazil}

\ead{mucio@cbpf.br}


\begin{abstract}
The study of multi-band superconductivity is relevant for a variety of systems, from ultra cold atoms with population imbalance to particle physics,  and
condensed matter. As a consequence, this problem has been widely investigated bringing to light many new and interesting phenomena.
In this work we point out and explore a correspondence between a two-band metal with a $k$-dependent hybridization and a uniformly polarized fermionic system in the presence
of spin-orbit coupling (SOC). We study the ground state phase diagram of the metal  in the presence of an attractive interaction.  We find  remarkable  superconducting properties whenever hybridization mixes orbitals of different parities in neighboring sites. We show that this mechanism enhances superconductivity and drives the crossover from weak to strong coupling in analogy with SOC in cold atoms. We obtain the quantum phase transitions  between the normal and superfluid states, as the intensity of different parameters characterizing the metal are varied, including Lifshitz  transitions, with no symmetry breaking, associated with the appearance of soft modes in the Fermi surface. 

\end{abstract}
\maketitle


\section{Introduction}

The development and progress of the  techniques to study ultra-cold atomic systems
has made them an ideal and clean platform to investigate condensed matter systems.
They allow to tune the relevant interactions in a large range and 
consequently to explore the  phase diagrams of these many-body systems.
More recently, the spin-orbit interaction has been implemented in cold atoms\cite{Y. J. Lin1, Y. J. Lin2} revealing still richer
phase diagrams. This interaction allows for quantum phase transitions which do not present
the usual symmetry-breaking phenomenon of the Landau paradigm and are best characterized in terms of topological transitions.

This work focus on the study of asymmetric superconductors~\cite{Nosso},  where different types of quasi-particles, the electrons arising from different orbitals, coexist at a common Fermi
surface~\cite{caldas}. These may also be atomic systems, with atoms in different nuclear states~\cite{nature} or colored superconductors, as found in the core of neutron stars~\cite{6,casalbuoni,7}, where the particles are different quarks.

A common parameter that characterizes asymmetric superfluids is the mismatch $\delta k_F$ between the Fermi wave-vectors associated with different quasi-particles.   The quantum phase diagram of these superfluids in the limit of very large mismatches where, even at $T=0$,  they are in the normal phase has been previously investigated~\cite{heron}.  As the mismatch is {\it reduced} they present an instability to an inhomogeneous superfluid state characterized by a space modulated order parameter, known as FFLO phase~\cite{fflo}. In this work we study the opposite limit of small mismatches where the ground state is a homogeneous superfluid.
We consider a two-band metal with inter-band attractive interactions and hybridization between them. In the case these bands are spin-up and down bands of a system polarized by a uniform  magnetic field, we show that under certain conditions, a $k$-dependent hybridization is formally equivalent to a spin-orbit interaction between the polarized bands.  

Our results on the influence of hybridization on superconductivity have remarkable implications. Whenever hybridization occurs  among orbitals with different parities, as $p-d$ or $d-f$ orbitals, we show that it enhances superconductivity and promotes a crossover from pure BCS to a strong coupling Bose-Einstein condensation (BEC) of pairs. The  case of $p-d$ hybridization is relevant for the  high-$T_c$ superconductors~\cite{pdhightc} and that of $d-f$ for heavy fermion materials~\cite{dfhf,coqblin,aline}. Since hybridization can be controlled by doping or pressure our results have exciting consequences for these systems.

The problem of superfluidity in the presence of spin orbit interaction has recently received a lot of attention~\cite{reviewSOC,3dSOC,3dSOCTri,sademello,sademelo2,3dSOC1,alicea,chineses,sato}. The general approach is to introduce the {\it helicity basis} in which the kinetic part of the Hamiltonian together with the Zeeman and spin-orbit terms are diagonal. Next a BCS decoupling is used  to deal with the many-body attractive interaction which is written in the  helicity basis~\cite{alicea}.
The superfluid order parameter  now contains  {\it triplet} and {\it singlet}  contributions arising from  pairing states with the same or opposite helicities, respectively~\cite{alicea,chineses,sato,shenoy,gorkov}. Here we  treat all terms of the Hamiltonian, that consist of the kinetic part, the hybridization or SOC, the  BCS decoupled attractive interaction and the Zeeman term on the same footing. This allows us to consider a single order parameter instead of several pairing amplitudes that arise in the helicity basis~\cite{alicea}.  Of course both methods should yield equivalent results, as we discuss below.
Furthermore, as a mathematical tool, instead of using generalized Bogoliubov transformations, we use Green's functions and the equations of motion method.

\section{Model and formalism}

We consider a model with two types of quasi-particles, $a$ and $b$, arising from different atomic orbitals with an
attractive inter-band interaction $g$,  and a hybridization term $V(k)=V_k$ that
mixes different quasi-particles states~\cite{Nosso,aline,coleman}. This one-body mixing term
$V_k$ is related to the overlap of the wave functions on the same or neighboring sites and can be tuned by
external parameters, like pressure or doping. The Hamiltonian is given by
\begin{eqnarray}
&  H=\sum\limits_{k\sigma}\epsilon_{k}^{a}a_{k\sigma}^{\dagger
}a_{k\sigma}+\sum\limits_{k\sigma}\epsilon_{k}^{b}b_{k\sigma}%
^{\dagger}b_{k\sigma}\nonumber\\
&\!  -\! g\!\!\sum\limits_{kk^{\prime}\sigma}a_{k\prime\sigma}^{\dagger
}b_{-k\prime-\sigma}^{\dagger}b_{-k-\sigma}a_{k\sigma}
\!+\! \sum\limits_{k\sigma}\!\left( V_{k} a_{k \sigma}^{\dagger
}b_{k\sigma}+V_{k}^{*}b_{k\sigma}^{\dagger}a_{k\sigma}\!\right)  \label{eq01}%
\end{eqnarray}
where $a_{k\sigma}^{\dagger}$ and $b_{k\sigma}^{\dagger}$ are creation
operators for the $a$ and $b$ quasi-particles, respectively and $g>0$.
The dispersion relations $\epsilon_{k}^{l}%
=\frac{\hbar^{2} k^{2}}{2m_{l}}-\mu_{l}$ ( $l=a,b$), where we allow for different masses and chemical potentials. We set $\hbar=1$.
The motivation for considering inter-band attractive interactions is that, as argued by many authors,  in heavy fermions the main contribution to superconductivity is due to hybrid or inter-orbital pairs involving f-electrons and conduction electrons, which arise from the dominant Kondo interaction~\cite{khomski}. For the copper oxides in some versions of the $d-p$ model~\cite{dpgeral}, it is argued that the $d-p$ interaction has a predominant role~\cite{dp}. 

An interesting feature of the above Hamiltonian is that if the bands $a$ and $b$ 
are taken as the up-spin and down-spin bands of a single band system polarized by an external magnetic field $h$, the hybridization
term now mixes different spin states. Then, depending on the symmetry properties of $V_k$, this problem becomes formally similar to that of a non-centrosymmetric system in the presence of Rashba spin-orbit interaction~\cite{3dSOC2} as we discuss below.  

Thus, Hamiltonian, Eq.~\ref{eq01} describes either a hybridized two-band system or a polarized single band material with  spin-dependent tunneling.  In both cases, there is an attractive interaction between the different quasi-particles. 
Notice that in spite of the formal similarity, spin degrees of freedom are important and distinguish the two-problems: the hybridization problem, which mixes fermions in different bands with the same spin and the spin-orbit Rashba interaction that mixes fermions in the same orbital band but with different spins. This distinction becomes most important in the presence of a magnetic field.

We will consider here the ground state phase diagram and topological properties of a 3d s-wave superfluid described by Eq.~\ref{eq01}.
The order parameter that characterizes the superfluid phase  is, $\Delta_{ab}=g\sum
\limits_{k\sigma}\left\langle b_{-k-\sigma}a_{k\sigma}\right\rangle $. 

Within the BCS approximation, Eq.~\ref{eq01} can be exactly diagonalized, either using a generalized Bogoliubov transformation or using the equations of motion for the Green's function~\cite{Nosso,tyablikov}. Here we use the latter method and obtain the anomalous correlation functions $\left\langle b_{-k-\sigma}a_{k\sigma}\right\rangle $  from the corresponding anomalous Greens function, $\left\langle \left\langle a_{k\sigma};b_{-k-\sigma}\right\rangle
\right\rangle_{\omega} $  \cite{gorkov,tyablikov}. The poles of the Green's function also yield the spectrum of
excitations in the superconducting phase. Excitonic types of correlations that simply renormalize the hybridization
\cite{sarasua} are neglected.  Finally, the anomalous frequency dependent propagator, from which 
the order parameter can be
self-consistently obtained, is given by \cite{Nosso}, \ \
\begin{equation}
\left\langle \left\langle a_{k\sigma};b_{-k-\sigma}\right\rangle \right\rangle_{\omega}
=\frac{\Delta_{ab}D_k(\omega)}{\omega^{4}+C_k \omega^{2}+ F_k}.
\label{eq02}%
\end{equation}
As we will see below, the values of the quantities $C_k$, $D_k$ and $F_k$ depend on a crucial manner in the symmetry properties of the hybridization $V_k$ under space inversion symmetry. We distinguish between two cases: symmetric hybridization, such that, $V(-k)=V(k)$ and anti-symmetric where $V(-k)=-V(k)$.  Anti-symmetric hybridization can occur when one mixes orbitals with angular momenta \textit{l} and \textit{l+1} in neighboring sites. This is the case of the $V_{df}$ hybridization between orbitals $d$ and $f$ in rare-earth and actinide based systems~\cite{coqblin} or $V_{pd}$ like in transition metals oxides~\cite{pdhightc}. Due to the different parities of the orbitals with orbital momenta \textit{l}  and \textit{l+1} the hybridization breaks inversion symmetry and it is odd in $k$. This occurs even for centro-symmetric systems, like a cubic lattice, where assuming, for example that {\bf k} is in the x-direction one gets~\cite{coqblin,sialso} $V_k \propto sink_x a$. The anti-symmetric hybridization,  does not mix states at the band edges $k=0$ and $\mathbf{k}=(\pi/a,\pi/a,\pi/a)$. In the former case it is similar to the Rashba spin-orbit coupling.

\section{Symmetric hybridization}


This is the case where $V(-k)=V(k)$. Using this property in the equations of motion, where terms of the type $V(-k)$ arise due to the BCS interaction that mixes states with opposite momenta, the anomalous frequency dependent propagator is given by Eq.~\ref{eq02} with~\cite{Nosso}, 
\[
D_k(\omega)=\Delta_{ab}^{2}\!-\!|V_k|^{2}\!-\!\left(  \omega\!-\!\epsilon_{k}%
^{b}\right)  \left(  \omega\!+\!\epsilon_{k}^{a}\right),   
\]
and
\begin{eqnarray}
C_k  &  =-\left[  \epsilon_{k}^{a2}+\epsilon_{k}^{b2}+2\left(
\Delta_{ab}^{2}+|V_k|^{2}\right)  \right], \label{eq04}\\
F_k &  =\left[  \epsilon_{k}^{a}\epsilon_{k}^{b}-\left(  |V_k|^{2}-\Delta
_{ab}^{2}\right)  \right]  ^{2}. \nonumber
\end{eqnarray}
The poles of the propagators yield the energies ($\omega_{1,2}(k)$ and $\omega_{3,4}(k)=-\omega_{1,2}(k)$) of the excitations in the superconducting phase. Also from the discontinuity of the Greens function, Eq.~\ref{eq02}, on the real
axis we can obtain the anomalous correlation function characterizing the
superconducting state. 

The condition for having excitations with zero energy is,
\begin{equation}
F_{k}=\left[  \epsilon_{k}^{a}\epsilon_{k}%
^{b}-\left(  |V_{k}|^{2}-\Delta_{ab}^{2} \right)  \right]^{2}=0.
\label{cond1}
\end{equation} 
For a constant hybridization $V_k=V_0$, this  occurs for $V_0=\Delta_{ab}$, in which case, gapless excitations appear at
$k=k_{F}^{a}$ and $k=k_{F}^{b}$, where $\epsilon_{k}^{a}=0$ and $\epsilon_{k}^{b}=0$. 

The
energy of the excitations obtained from the poles of Eq.~\ref{eq02} are given by, 
\begin{equation}
\omega_{1,2}(k)=\sqrt{A_{k}\pm\sqrt{B_{k}}} \label{dispersion}%
\end{equation}
with,
\begin{equation}
A_{k}=\frac{\epsilon_{k}^{a2}+\epsilon_{k}^{b2}}{2}+\Delta_{ab}^{2}%
+|V_k|^{2} \label{eq05}%
\end{equation}
and%
\begin{eqnarray}
\label{eq06}
B_{k}  &  =\left(  \frac{\epsilon_{k}^{a2}-\epsilon_{k}^{b2}}{2}\right)
^{2}+|V_k|^{2}\left(  \epsilon_{k}^{a}+\epsilon_{k}^{b}\right)  ^{2}+\Delta
_{ab}^{2}\left(  \epsilon_{k}^{a}-\epsilon_{k}^{b}\right)  ^{2} +4|V_k|^{2}\Delta_{ab}^{2}. 
\end{eqnarray}

%
%

\subsection{Two-band system with hybridization}

Let us apply these results for a two-band superconductor in zero external magnetic field with the ratio of the quasi-particles masses given by, $m_a/m_b=\alpha$. For simplicity we assume that the dispersion relations of these bands are given by,  $\epsilon_k^b=\alpha \epsilon_k^a =\alpha \epsilon_k$. The condition for the existence of zero energy modes is  given by,
\begin{equation}
  \alpha \epsilon_{k}^2-\left(  |V_k|^{2}-\Delta_{ab}^{2}\right)  =0.
\end{equation}
This equation can be conveniently normalized and rewritten as:
\begin{equation}
\label{conda}
  \alpha (\tilde{k}_z^2+\tilde{k}_{\perp}^2 -1)^2-\left(  |\tilde{V}_k|^{2}-\tilde{\Delta}_{ab}^{2}\right)  =0.
\end{equation}
where $\tilde{k}=k/k_F$, $\tilde{V_k}=V_k/E_F$, $\tilde{\Delta}_{ab}=\Delta_{ab}/E_F$, where $k_F$ and $E_F=k_F^2/2m_a$ are the Fermi wave vector and Fermi energy of the unhybridized system, respectively. Also $k_{\perp}=\sqrt{k_x^2+k_y^2}$.

As pointed out before, for a constant hybridization,  Eq.~\ref{conda} is satisfied for $V=\Delta_{ab}$ and $k_z=k_{\perp}=k_F$. In this case, when hybridization  increases from zero there is a discontinuous quantum first order phase transition from the superconductor to the normal state as it reaches the critical value  $V_c=\Delta_{ab}$. This is associated with an instability of the whole Fermi surface of the system with respect to zero energy excitations.

\begin{figure}[th]
\centering{\includegraphics[scale=0.8]{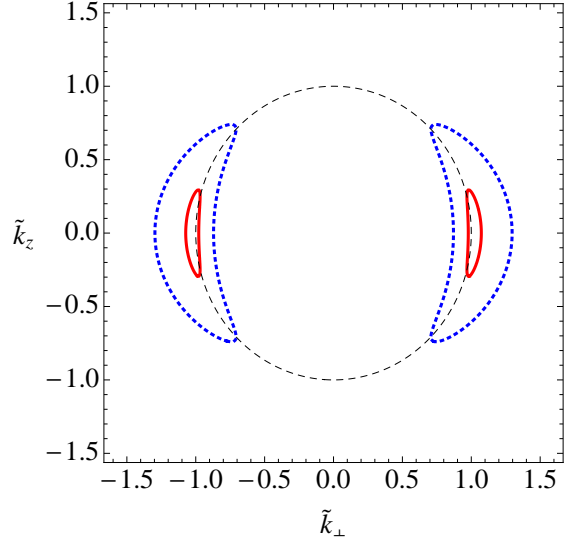}}\caption{ (Color online) Contour plots of Eq.~\ref{conda} for $\alpha=0.5$, $\tilde{\Delta}_{ab}=0.15$ and $\tilde{\beta}=0.3$ (dotted) and $\tilde{\beta}=0.16 >\tilde{\beta}_c=\tilde{\Delta}_{ab}$ (thick line).  The dashed line is projection of the Fermi surface of the unhybridized system. }%
\label{fig1}%
\end{figure}

In real systems in many cases mixing occurs among orbitals of different sites and the k-dependence of the  hybridization must be taken into account. Let us consider the case of YbAlB$_2$, where mixing occurs mainly in a plane~\cite{aline} and can be modeled by  $V_k=\beta k_{\perp}^2$. Substituting this expression for $V_k$  in Eq.~\ref{conda}, we see that the condition for zero modes is now quite different from the constant $V$ case. For $\beta k_F^2< \Delta_{ab}$ the system is a standard superconductor with gaped excitations. However, at $\tilde{\beta}_c=\tilde{\Delta}_{ab}$,  where we defined $\tilde{\beta}=\beta k_F^2/E_F$, there is a zero temperature phase transition to a  superconducting state with a line of zero energy excitations at the Fermi surface of the unhybridized system. This line  occurs for $\tilde{k}_{\perp}=1$, $\tilde{k}_z=0$. As $\tilde{\beta}$ increases this line splits in two, one in each hemisphere of the Fermi surface, as shown in Fig.~\ref{fig1}.
This quantum phase transition (QPT) occurs without any symmetry breaking, since as shown below, the system remains a superfluid for $\tilde{\beta}>\tilde{\beta}_c=\tilde{\Delta}_{ab}$. This transition is a {\it Lifshitz transition} and the associated quantum critical exponents are well known~\cite{volovik,livro}.

\begin{figure}[th]
\centering{\includegraphics[scale=0.8]{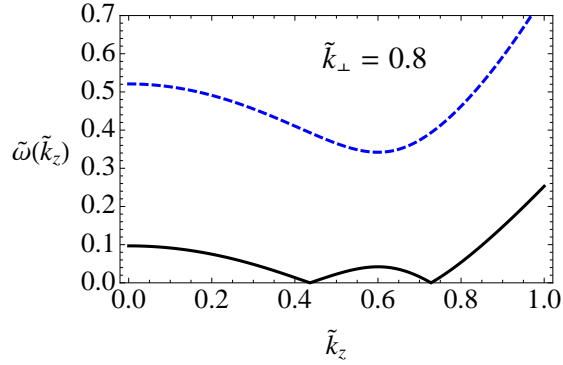}}\caption{(Color
online) Dispersion relations as functions of $\tilde{k}_z$ for a fixed value of $\tilde{k}_{\perp}$ for $\alpha=0.5$, $\tilde{\Delta}_{ab}=0.15$ and $\tilde{\beta}=0.3>\tilde{\beta}_c=\tilde{\Delta}_{ab}$. For this value of $\tilde{k}_{\perp}$ the excitations  are gapless at two values of $\tilde{k}_z$. }%
\label{fig2}%
\end{figure}

Figure~\ref{fig1} shows  surfaces of zero energy modes for $\tilde{\beta}>\tilde{\beta}_c=\tilde{\Delta}_{ab}$. These surfaces cross the original Fermi surface at two circles, one in each hemisphere, where the energy of the excitations vanishes, as shown in this figure.  Figure~\ref{fig2} shows the dispersion relations of the excitations for a fixed value of $k_{\perp}$, using the parameters of Fig.~\ref{fig1} with $\tilde{\beta}=0.4 > \tilde{\beta}_c$. For the chosen value of $\tilde{k}_{\perp}$ there are two circles with zero energy modes.

It is important to emphasize that superconductivity survives the Lifshitz transition, at least at zero temperature.  This can be verified using the self-consistent gap equation to calculate the superconducting order parameter.  This equation can be written as, 
\begin{eqnarray}
\label{gapV}
\frac{1}{\rho g}&=\frac{V}{(2 \pi)^3} 4 \pi \Big\{\! \int_0^{k_F+\delta}\!\!\!dk_z \int_0^{\sqrt{(k_F+\delta)^2-k_z^2}}\!\!dk_{\perp}k_{\perp} f_a(k_z, k_{\perp})  \nonumber \\
&\!-\int_0^{k_F-\delta}\!\!\!dk_z \int_0^{\sqrt{(k_F-\delta)^2-k_z^2}}\!\!dk_{\perp}k_{\perp} f_a(k_z, k_{\perp})\!\Big\}, \nonumber \\
\end{eqnarray}

where

\begin{equation}
f_a(k_z, k_{\perp})\!=\!\frac{1}{4 \pi (\omega_1+ \omega_2)}\!\left[\tanh(\frac{\omega_1}{2 k_B T})\!+\! \tanh(\frac{ \omega_2}{2 k_B T})\right].
\label{eq07}%
\end{equation}
The energies $\omega_i$ above are given by Eq.~\ref{dispersion}  and are functions of $\epsilon_k$ and $\alpha$, since we are using homothetic bands, such that,  $\epsilon_{k}^{a}=\epsilon_k$ and $\epsilon_{k}^{b}=\alpha \epsilon_k$. Furthermore, 
$$\epsilon_k = E_F(\tilde{k}_z^2 + \tilde{k}_{\perp}^2 -1).$$

\begin{figure}[th]
\centering{\includegraphics[scale=0.8]{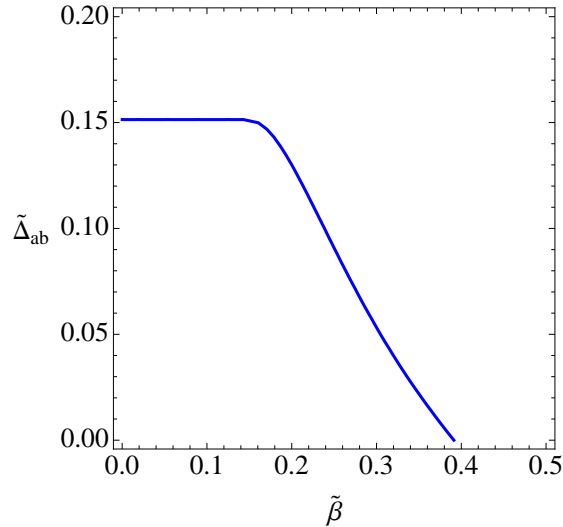}}\caption{(Color
online) The superfluid order parameter as a function of the strength of hybridization $\tilde{\beta}$. The transition to a topological superconductor occurs for $\tilde{\beta}_c=\tilde{\Delta}_{ab}=0.15$. We used $\alpha=0.5$, $\rho g = 0.25$ and the cutoff (renormalized by $k_F$), $\delta=0.05$. }%
\label{fig3}%
\end{figure}

We solve the gap equation, Eq.~\ref{gapV}, as a function of $\tilde{\beta}$ at zero temperature.
As shown in Fig.~\ref{fig3},  the order parameter $\Delta_{ab}$ remains finite even for $\tilde{\beta} > \tilde{\beta}_c$.
However, $\Delta_{ab}$ is sensitive to the Lifshitz transition and for sufficiently large $\tilde{\beta}> \tilde{\beta}_c$ superfluidity is eventually  destroyed continuously at a quantum critical point~\cite{livro}.

\subsection{Weak to strong coupling crossover}

In case the attractive interaction becomes sufficiently strong, we have to solve self-consistently the number and  gap equations to obtain the chemical potential and the order parameter. We consider the two-band case and as usual, when dealing with the strong coupling limit,  we introduce the scattering length $a_s$  as a convenient   renormalization that allows to eliminate the ultraviolet divergence in the gap equation. This can then be written as:
\begin{equation}\label{gapbecbcs}
- \frac{m}{4 \pi a_s} = \sum_k \left(\frac{1}{\omega_1 + \omega_2} - \frac{1}{(1+\alpha) \epsilon_k}\right).
\end{equation}
The energies $\omega_{1,2}(k)$ are given by Eqs.~\ref{dispersion} and as before we use the homothetic relations, $\epsilon_k^b=\alpha \epsilon_k^a= \alpha \epsilon_k$.
The number equation is given by,
\begin{equation}\label{mubecbcs}
N = \sum_k \left(1- \frac{(1+\alpha) \epsilon_k}{\omega_1 +\omega_2}\right),
\end{equation}
where $N=N_a + N_b$ is the total number  of electrons  in the two bands. Equations~\ref{gapbecbcs} and \ref{mubecbcs} determine the gap and the chemical potential of the two-band system.
The calculations are implemented substituting the sums by integrals, $ \sum_k \rightarrow (1/2 \pi^2) \int dk_z \int dk_{\perp} k_{\perp}$, where we took a unitary volume, with  the limits of the integrals extending to $\infty$ since they now converge because the integrands vanish in this limit.  
\begin{figure}[th]
\centering{\includegraphics[scale=0.85]{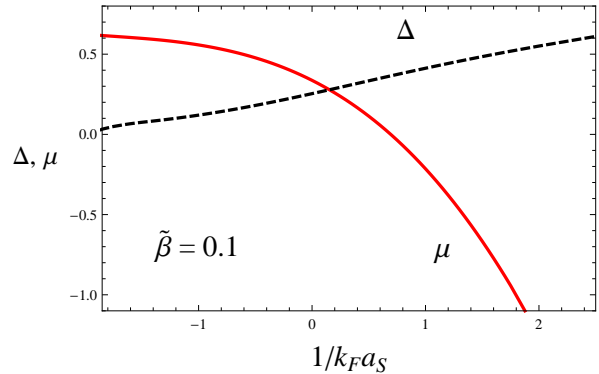}}
\caption{Zero temperature order parameter $\Delta=\tilde{\Delta}_{ab}$ and chemical potential as  functions of the ratio $1/k_F a_s$ for the case the hybridization $\tilde{V}= \tilde{\beta} k_{\perp}$ with $\tilde{\beta}=0.1$. The ratio of the masses of the quasi-particles  is taken as $\alpha=0.1$. }
\label{fig4}
\end{figure}
\begin{figure}[h]
\centering{\includegraphics[scale=0.85]{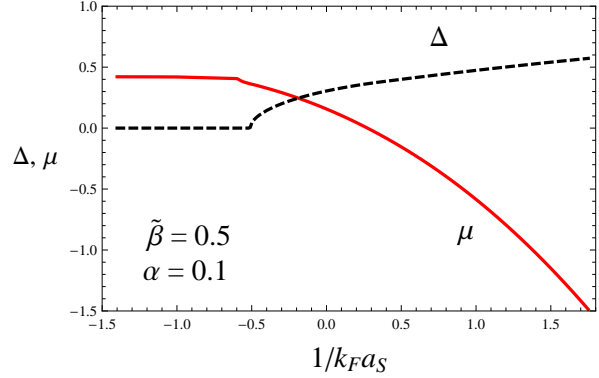}}\caption{ Zero temperature order parameter $\Delta=\tilde{\Delta}_{ab}$ and chemical potential as  functions of the ratio $1/k_F a_s$ for $\tilde{V}= \tilde{\beta} k_{\perp}$ with $\tilde{\beta}=0.5$. The ratio of the masses of the quasi-particles  is taken as $\alpha=0.1$. The crossover to the BEC regime occurs for smaller ratios $1/k_F a_s$  as the hybridization strength $\tilde{\beta}$ increases.  Notice the existence of a superconducting quantum critical point (SQCP) at a minimum critical value of the coupling $1/k_F a_s$ for superconductivity to appear.}
\label{fig5}
\end{figure}
\begin{figure}[h]
\centering{\includegraphics[scale=0.85]{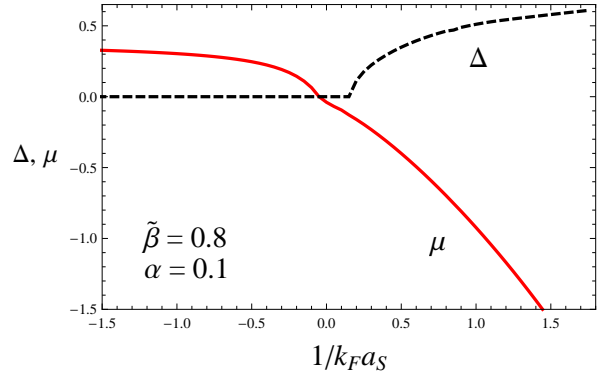}}\caption{ Zero temperature order parameter $\Delta=\tilde{\Delta}_{ab}$ and chemical potential as  functions of the ratio $1/k_F a_s$ for $\tilde{V}= \tilde{\beta} k_{\perp}$ with $\tilde{\beta}=0.8$.  The ratio of the masses of the quasi-particles  is taken as $\alpha=0.1$. As in Fig.~\ref{fig5} there is  a SQCP which in this case appears in the BEC regime where the chemical potential is negative.  }
\label{fig6}
\end{figure}
In Figs.~\ref{fig4}, \ref{fig5} and~\ref{fig6}, we show, the gap and the chemical potential as  functions of $1/k_F a_s$ for different values of the hybridization strength $\tilde{\beta}$. For easier convergence of the integrals we used the form of the hybridization $\tilde{V}=\tilde{\beta} k_{\perp}$, varying linear with $k_{\perp}$ as in the SOC problem.   As  $\tilde{\beta}$ increases, it becomes necessary a minimum value of the attractive interaction for superconductivity to be stabilized in the system. This is in agreement with the weak coupling results that have shown the deleterious effect of the symmetric hybridization in superconductivity. Then it is natural to expect that above a critical value of $\tilde{\beta}$, a minimum value for the attractive interaction is required to stabilize superconductivity. In our case this is clearly associated with the presence of a quantum critical point at a critical value of the coupling $(1/k_F a_s)$.  Notice that if $\tilde{\beta}$ is sufficiently large ($\tilde{\beta}=0.8$) superconductivity appears already in the BEC region where the chemical potential is negative, as shown in Fig.~\ref{fig6}.

\subsection{Lifshitz transitions in a polarized single band system}

Let us now consider the case  $a$ and $b$ are  up and down spin-bands with the degeneracy raised by an external longitudinal magnetic field $h$, such that,
\begin{eqnarray}
\label{dispfield}
\epsilon_k^a= \epsilon_k +h \nonumber \\
\epsilon_k^b= \epsilon_k - h,
\end{eqnarray}
where $\epsilon_k=k^2/2m-\mu$.  {\it Hybridization} now mixes different spin bands, but in the symmetric case, i.e., with $V(-k)=V(k)$, it does not correspond to any real physical interaction in a polarized single band system. As  it turns out  to be interesting to study this case, we can imagine it arises from an external $k$-dependent {\it transverse magnetic field} $h_x(k)=V(k)=\gamma k_{\perp}$ applied in the $x$-direction, besides the longitudinal uniform Zeeman magnetic field $h$.   Furthermore, since further down we consider the anti-symmetric case that corresponds to a Rashba spin-orbit coupling, comparing the two cases will show the profound influence the symmetry properties of $V(k)$ have on the phase diagram of the system.

Substituting Eqs.~\ref{dispfield} in the expressions for the energy of the quasi-particle excitations, Eqs.~\ref{dispersion}, these simplify considerably and we get,
\begin{equation}
\omega_{1,2}(k)= E_k \pm \sqrt{|V_k|^2 + h^2},
\label{hdispersion}%
\end{equation}
where $E_k=\sqrt{\epsilon_k^2 +\Delta_{ab}^2}$ and $|V_k|^2=\gamma^2 k_{\perp}^2$ with $\gamma$ the intensity of the transverse field (we use $\gamma$ instead of $\beta$ to distinguish  from conventional hybridization).
The condition for having zero energy modes, $F_k=0$ (see Eq.~\ref{eq04})  is now given by,
\begin{equation}
\epsilon_k^2-h^2-\gamma^2 k_{\perp}^2+\Delta_{ab}^2=0.
\end{equation}
This equation can be written in the form,
\begin{equation}
\label{fund}
\left( \tilde{k}_z^2+\tilde{k}_{\perp}^2 -1\right)^2 - \tilde{h}^2 -\tilde{\gamma}^2 \tilde{k}_{\perp}^2 +\tilde{\Delta}_{ab}^2=0,
\end{equation}
where $\tilde{k}=k/k_F$, $\tilde{h}=h/E_F$, $\tilde{\Delta}_{ab}=\Delta_{ab}/E_F$ and $\tilde{\gamma}=(\gamma k_F)/E_F$, where $E_F$ is the Fermi energy.

Defining the functions,
\begin{equation}
\mathcal{F}^{\pm}= 4 \pi \int_0^{1 \pm \delta} d  \tilde{k}_z  \int_0^{\sqrt{(1 \pm \delta)^2 - \tilde{k}_z^2}} \frac{d \tilde{k}_{\perp} \tilde{k}_{\perp} }{(\tilde{\omega}_1+|\tilde{\omega}_2|)},
\end{equation}
where $\tilde{\omega}_{i}=\omega_{i}/E_F$,  with $\omega_{i}$ given by Eqs.~\ref{hdispersion}, and $\delta$ is a momentum  cutoff (normalized by $k_F$),  the gap equation can be cast in the form,
\begin{eqnarray}
\frac{1}{g \rho}=\mathcal{F}^{+} - \mathcal{F}^{-}.
\end{eqnarray}

\begin{figure}[th]
\centering{\includegraphics[scale=0.8]{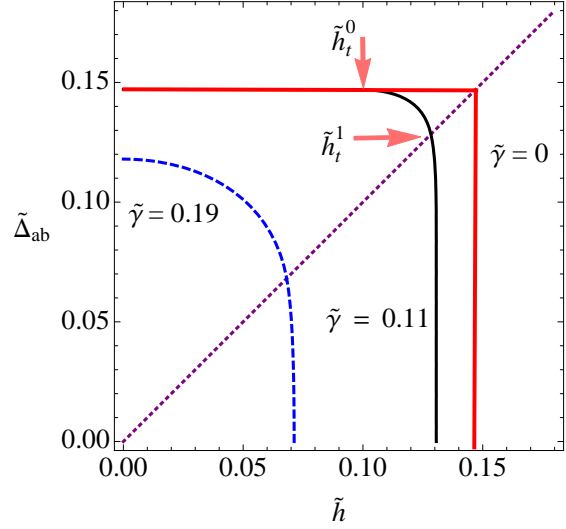}}\caption{(Color online) The zero temperature order parameter $\Delta=\tilde{\Delta}_{ab}$ as a function of the external longitudinal Zeeman magnetic field for different values of the transverse field.  The dotted line marks the Lifshitz transition at which Fermi points appear on the Fermi surface.  The arrows point the fields for which the Lifshitz transitions occur for the case $\tilde{\gamma}=0.11$. We have used as numerical parameters, $g \rho = 0.25$ and $\delta=0.05$. }%
\label{fig7}%
\end{figure}

In the  absence of the transverse field, $\gamma=0$, the order parameter is constant up to a critical longitudinal field $h_c=\Delta_{ab}$, at which there is a first order quantum phase transition to the normal state where the order parameter $\Delta_{ab}$ vanishes abruptly, as shown in Fig.~\ref{fig7}. This instability is associated with the appearance of  zero energy modes at the whole Fermi surface  of the non-polarized system. This is a Lifshitz transition, in this case associated with a broken symmetry since it is accompanied by the disappearance of superconductivity. 

For $\gamma \ne 0$, the superconducting phase is also destroyed by the longitudinal magnetic field, but the transition instead of being abrupt becomes rounded due to the transverse field. We have to distinguish between two cases, $\tilde{\gamma} < \tilde{\Delta}_{ab}^{0}$ and $ \tilde{\gamma} > \tilde{\Delta}_{ab}^{0} $ where $\tilde{\Delta}_{ab}^{0} = \tilde{\Delta}_{ab}(\tilde{h}=0,\tilde{\gamma}=0)$.

In the case $\tilde{\gamma} \le \tilde{\Delta}_{ab}^{0}$, the first Lifshitz transition occurs for $\tilde{h}_t^0=\sqrt{ (\tilde{\Delta}_{ab}^{0})^2-\tilde{\gamma}^2}$, where using the numerical values of the parameters in Fig.~\ref{fig7},  $ \tilde{\Delta}_{ab}^{0}=0.148$. This Lifshitz transition is associated with the appearance of a line of zero energy modes in the equator of the original Fermi surface, at $k_{\perp}=1$, $k_z=0$. In Fig.~\ref{fig4} for  $\tilde{\gamma}=0.11$ and using the value of $\tilde{\Delta}_{ab}^{0}$  above, this occurs for $\tilde{h}_t^0 \approx 0.1$. As the field further increases, the line of zero modes splits in two, one in each hemisphere of the Fermi surface (see Fig.~\ref{fig1}) and finally at $h_t^{1}=\Delta_{ab}(h_t^{1})$ these shrink to points in the poles of the spherical Fermi surface.

The case $\tilde{\gamma}=\tilde{\Delta}_{ab}^{0}$ is in its own class since, as can be seen from Eq.~\ref{fund}, for $h=0$ there is a collapse of the whole Fermi surface $(k_z^2+k_{\perp}^2-1=0)$ for $\tilde{\gamma}=\tilde{\Delta}_{ab}^{0}$. In this case superconductivity disappears abruptly in zero external longitudinal field. This first order transition is similar to that which occurs for $\tilde{\gamma}=0$ and $\tilde{h}=\tilde{\Delta}_{ab}^{0}$, as shown in Fig.~\ref{fig7}.

As $\tilde{\gamma}$ increases beyond $\tilde{\Delta}_{ab}^{0}$, i.e., for $\tilde{\gamma}>\tilde{\Delta}_{ab}^{0} = 0.148$, using the parameters of Fig.~\ref{fig7},
the Lifshitz transition now occurs exclusively for 
$h_t^{1}=\Delta_{ab}(h_{t}^{1})$ and is related to the appearance of Fermi points at $k_{\perp}=0$, $k_z=\pm1$, i.e.,  on the poles of the original Fermi sphere. The line of these transitions is also shown in Fig.~\ref{fig7} (straight dotted line).
Notice that these transitions occur without necessarily destroying the superconducting phase at least at $T=0$.
It is worth pointing out, as can be seen from Eq.~\ref{fund}, that a zero energy mode can also appear at $k=0$, for a field $\tilde{h}=\sqrt{1+\tilde{\Delta}_{ab}^2}$. This field is much larger than $\tilde{h}_{t}^{0}$    and $\tilde{h}_{t}^{1}$ considered previously and for reasonable values of the other parameters (smaller than 1), superconductivity has been already destroyed before $h$ reaches this value.

\begin{figure}[th]
\centering{\includegraphics[scale=0.8]{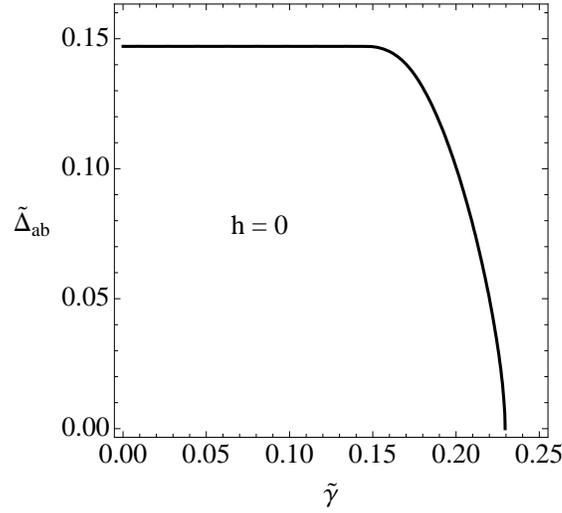}}\caption{ The zero temperature order parameter $\Delta=\tilde{\Delta}_{ab}$ as a function of the transverse field for zero external longitudinal Zeeman magnetic field. We used the same parameters as in Fig.~\ref{fig7}. }%
\label{fig8}%
\end{figure}

In order to obtain a complete picture of the influence of the transverse field in the phase diagram, we show in Fig.~\ref{fig8} the effect of this field on superfluidity. For zero external longitudinal magnetic field there is a critical value of the transverse field $\gamma_c$ for which superconductivity disappears. Using the same numerical parameters as in Fig.~\ref{fig7}, we obtain $\tilde{\gamma}_c=0.23$ as shown in Fig.~\ref{fig8}. 


Let us consider the dispersion relation of the modes which soften as $\tilde{h} \rightarrow \tilde{h}_t$ at $k_{\perp}=1$, $k_z=0$ as for the case $\tilde{\gamma} \le \tilde{\Delta}_{ab}^{0}$.  This is given by,
\begin{equation}
\label{soft}
\omega(k_{\perp})=( \tilde{h}_t - \tilde{h})+ 2 \frac{(k_{\perp}-1)^2}{\tilde{\Delta}_{ab}^{0}}.
\end{equation}
where $\tilde{h}_t^1=\sqrt{(\tilde{\Delta}_{ab}^{0})^{2}-\tilde{\gamma}^2}$. This expansion is possible since the order parameter $\tilde{\Delta}_{ab}$ remains finite at the Lifshitz transition. The gap vanishes linearly close to this transition with a characteristic  exponent~\cite{livro} $\nu z=1$, while the spectrum in this case is {\it quadratic} in momentum.


\section{Non-symetric hybridization}

This is the case $V(-k)=-V(k)$. This situation may arise in non-centrosymmetric lattices but more interesting this occurs also in  symmetric lattices, if we consider hybridization among orbitals with opposite parities in neighboring sites, such as, $p-d$  or
$d-f$ hybridization that mixes orbitals with angular momentum~\cite{coqblin} $ l$ and $ l+1$. The former is relevant for the high-$T_c$ oxides and the latter for heavy fermion materials and actinide metals in general~\cite{coqblin}. Furthermore, many of the most interesting heavy fermion systems have tetragonal structures with rare-earths and transition metals in the planes perpendicular to $c$-axis, such that,  $d-f$ hybridization occurs predominantly in this plane. Also additional effects due to crystal fields may constrain mixing to take place mostly in the $ab$ plane~\cite{aline}. 


Using that $V(-k)=-V(k)$ in the equations of motion for the Green's functions, we find that the quantities, $C_k$, $D_k$ and $F_k$ in Eq.~\ref{eq04} are modified and the energy of the excitations in the superconducting phase are now given by,
$\pm \omega_{1,2}(k)$, where,
$$\tilde{\omega}_{1,2}(k)=\sqrt{\tilde{A}_k \pm \sqrt{\tilde{B}_k}}$$
with
\begin{equation}
\label{akt}
\tilde{A}_k=\frac{ \epsilon_k^{a2}+ \epsilon_k^{b2}}{2}
+\Delta_{ab}^2+|V_k|^2
   \end{equation}
   and
\begin{eqnarray}
\tilde{B}_{k}\!=\!\left(  \frac{\epsilon_{k}^{a2}-\epsilon_{k}^{b2}}{2}\right)^{2}\!\!\!+\!|V_k|^{2}\left(  \epsilon_{k}^{a}\!+\!\epsilon_{k}^{b}\right)^{2}+\Delta_{ab}^{2}\left(  \epsilon_{k}^{a}-\epsilon_{k}^{b}\right)^{2}\label{bkt}.
\end{eqnarray}
   
The condition for zero energy modes now is given by,
\begin{equation}
\left(\epsilon_k^{a} \epsilon_k^{b}+\Delta_{ab}^2-|V_k|^2\right)^2+4 \Delta_{ab}^2 |V_k|^2=0.
\end{equation}
Since this condition for any given $V_k$  can not be satisfied, there is no Lifshitz transition in this case. The situation is quite different in the presence of a longitudinal  external magnetic field where a pair of Fermi points appears at a Lifshitz transition~\cite{aseguir}.

The gap equation is also modified by the anti-symmetry property of $V(k)$.
It is now given by:

\begin{eqnarray}
&\frac{1}{ g}\! =\! \sum_{k} \!\frac{1}{4 \sqrt{\tilde{B}_k} } \Bigg\{ \! \frac{(\epsilon_k^{a} - \epsilon_k^b)^2}{2}  \left[\frac{\tanh( \beta \tilde{\omega}_1/2)}{\tilde{\omega}_1}\!-\! \frac{\tanh( \beta \tilde{\omega}_2/2)}{\tilde{\omega}_2} \right]+  \nonumber \\
&+ \sqrt{\tilde{B}_k} \left[\frac{\tanh( \beta \tilde{\omega}_1/2)}{\tilde{\omega}_1}\!+\! \frac{\tanh( \beta \tilde{\omega}_2/2)}{\tilde{\omega}_2} \right] \Bigg\}
\end{eqnarray}
and the number equation (at $T=0$):
\begin{eqnarray}
&N\! =\! \sum_{k} \Bigg\{ 1 - \frac{\epsilon_k^a+ \epsilon_k^b}{2(\tilde{\omega}_1 +\tilde{\omega}_2)}\left[ \frac{ \tilde{\omega}_1 \tilde{\omega}_2+ \varphi_k}{    \tilde{\omega}_1 \tilde{\omega}_2  } \right] \Bigg\}
\end{eqnarray}
where $\varphi_k= \epsilon_k^a \epsilon_k^b + \Delta^2 - |V_k|^2$ and $N=N_a+N_b$.

\begin{figure}[th]
\centering{\includegraphics[scale=0.8]{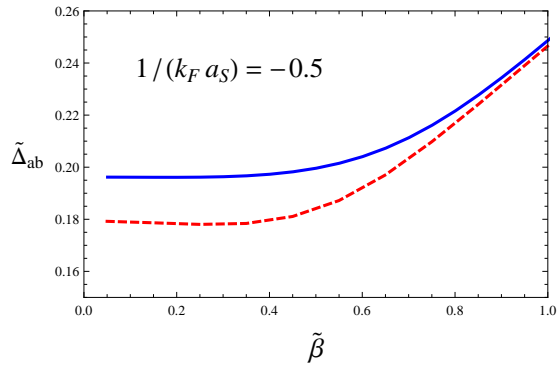}}\caption{(Color online) The zero temperature order parameter $\tilde{\Delta}_{ab}$ as a function of the strength of the hybridization for two mass ratios, $\alpha=0.5$ (full line) and $\alpha=0.25$ (dashed line).  }%
\label{fig9}%
\end{figure}
\begin{figure}[th]
\centering{\includegraphics[scale=0.8]{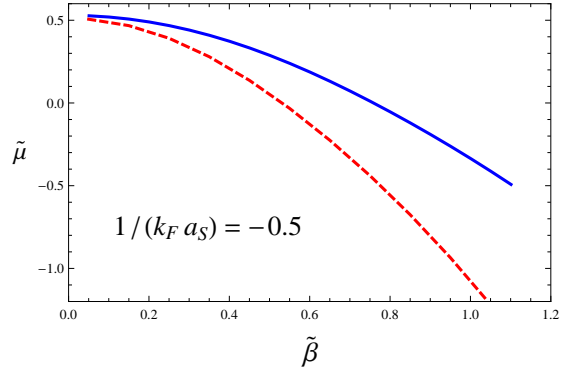}}\caption{(Color online) The zero temperature chemical potential $\tilde{\mu}$ as a function of the strength of the hybridization for two mass ratios, $\alpha=0.5$ (full line) and $\alpha=0.25$ (dashed line).  As hybridization increases and $\tilde{\mu}$ becomes negative, the main mechanism of superconductivity is the condensation of hybridons (see text).}%
\label{fig10}%
\end{figure}

Introducing the scattering length, as before, we solve self-consistently the equations above at zero temperature to obtain results for the superconducting gap and the chemical potential in the case of nearly two-dimensional systems where hybridization occurs mostly in a plane. Furthermore we take the functional form $|V(k)| = \tilde{\beta} k_{\perp}$ similar to the Rashba coupling. This is actually the form of $|V_{dp}(k)|$ for the square lattice of the $CuO_2$ planes in the tight-binding approximation~\cite{sialso} and in the limit of small $\mathbf{k}_{\perp}$. For simplicity we use the homothetic relations, $\epsilon_k^b=\alpha \epsilon_k^a$ with $\epsilon_k^a=\epsilon_k=k^2/2m - \mu$.

A remarkable result is shown in Figs.~\ref{fig9} and ~\ref{fig10} where we plot the gap and the chemical potential, for a mass ratio $\alpha=0.5$ and $1/(k_F a_S)=-0.5$ as functions of the intensity of the hybridization $\tilde{\beta}$ ($|V(k)| = \tilde{\beta} k_{\perp}$). Differently from the previous case of symmetric hybridization, as $\tilde{\beta}$ increases superconductivity is enhanced as indicated by the increase with $\tilde{\beta}$ of the renormalized gap $\tilde{\Delta}_{ab}$. Furthermore, as $\tilde{\beta}$ increases the chemical potential drops and becomes negative signaling a change of regime from BCS superconductivity to Bose-Einstein condensation of pairs.
Notice that this occurs for a value of the interaction $1/(k_F a_S)=-0.5$ which is typical of the weak-coupling BCS regime~\cite{leggett}.
This behavior had been noted previously in the context of atomic systems with spin-orbit interactions~\cite{shenoy} due to the formation of bound states by the Rashba SOC. In the context of condensed matter physics, this phenomenon, that we call the formation of {\it hybridons},  acquires a new significance due to the sensitivity of hybridization to doping and external pressure in these systems. Then, since hybridization can be tuned by external parameters, increasing $V_{pd}$ or $V_{df}$  provides a mechanism not only for increasing the critical temperatures in this type of superconductors but also to drive the BCS-BEC crossover.  
The class of materials with tetragonal structures for which non-symmetric hybridization occurs, namely the high-$T_c$ oxides~\cite{pdhightc} with $V_{dp}(k)$ and many heavy fermions~\cite{coqblin,aline} with $V_{df}(k)$ are of great interest.

Finally, notice that the two-band problem with asymmetric hybridization with $V(k)=\tilde{\beta} k_{\perp}$ maps exactly in the problem of a polarized single band system ($\alpha=1$) with Rashba SOC, both in the presence of attractive interactions.

In spite of the formal similarities pointed above between the odd parity hybridization case and the spin-orbit problem where the two species of fermions are labeled by spin, there is an important difference between these two problems. In the latter when one diagonalizes the kinetic energy and the spin-orbit coupling, the new quasi-particle operators involve a linear combination of creation and annihilation operators of electrons with  {\it different spins}~\cite{kampf}. In the former, the new quasi-particle operators that diagonalize the kinetic energy terms plus the hybridization involve also a linear combination of the original band operators but  {\it with the same spin}. A direct consequence of this difference is that in the spin-orbit problem, when the interaction terms of the BCS mean field Hamiltonian, $H_{BCS}= - \Delta_{ab} \sum_k (a^{\dagger}_{k \sigma} b^{\dagger}_{-k -\sigma} + b_{-k -\sigma} a_{k \sigma})$ are written in terms of the new quasi-particle operators, triplet correlations immediately arise. This is not the case in the mixing problem. If the inter-band interaction acts only in the s-wave channel it continues to do so in the new basis of hybridized states, even though anomalous induced correlations with a p-wave character can arise, as we discuss below.


A final comment concerns the role of self-energy corrections to the problem above. At finite temperatures the correct Bose-Einstein condensation temperature is obtained in the strong coupling limit of the BCS-BEC crossover only if one goes beyond mean-field and includes the self-energy, which enters in the calculation by considering fluctuations corrections \cite{randeria}. These corrections affect even the zero temperature behavior but in a quantitative way. However, the main point here is that  we showed that even for a {\it fixed weak-coupling interaction}, where fluctuations are negligible, the BCS-BEC  crossover can be reached by varying the strength of hybridization, as shown in Fig.~\ref{fig9}.

\section{Intra-band interaction}

Here we mention briefly the  case with attractive intra-band interactions, in the narrow $b$-band only, and for odd parity hybridization. The condition for zero modes is given by~\cite{Nosso,aseguir},
$$(\epsilon_k^a \epsilon_k^b - |V_k|^2)^2 + {\epsilon_k^a}^2 \Delta_{bb}^2=0,$$
where $\Delta_{bb}$ is the superconducting order parameter in our notation. Then, in this case there are no zero modes unless the hybridization vanishes at the Fermi wave-vector of the $a$-band. It turns out from the calculations that the energy of the excitations in the superconducting phase for symmetric and anti-symmetric (odd parity) hybridizations are formally the same and differ only by the specific functional form of $|V(k)|^2$. Furthermore, considering just the intra-band attractive interaction and a hybridization term, we find induced inter-band pairing correlations due to the hybridization in the form~\cite{aseguir}, $\Delta_{ab}(k) \propto V(k) \Delta_{bb}$. Then, for odd parity $V(k)$ the induced  inter-orbital pairing is of the $p$-wave type. A reverse effect occurs in the inter-band case treated here, but with $\Delta_{bb}(k) \propto V(k) \Delta_{ab}$~\cite{aseguir}. Notice that induced gaps do not appear in the zero mode equations. 

\section{Comparison with other approaches}

Instead of diagonalizing the full Hamiltonian, Eq.~\ref{eq01}, with the attractive interaction treated in the BCS approximation, it is a common approach in the literature \cite{3dSOC,3dSOC1,alicea,3dSOC2} to use the helicity basis and write the attractive interaction in this basis. The helicity basis is that which diagonalizes the part of the Hamiltonian containing the kinetic energy, the Rashba coupling and the Zeeman term. This has as eigenvalues \cite{alicea},
\begin{equation}
\epsilon^{\pm}_k=\epsilon_k \pm \sqrt{h^2+|V_k|^2},
\end{equation}
where $\pm$ refer to helicity states. If one uses a BCS approximation and writes the attractive interaction in this helicity basis, the energies of the quasi-particles in the superconducting state are obtained as~\cite{alicea,3dSOC2},
\begin{equation}
\label{dispgeral}
\Omega^{\pm}(k)= \sqrt{\left( E_k \pm \sqrt{|V_k|^2 + h^2} \right)^2+ |\Delta_{++}|^2},
\end{equation}
where $E_k=\sqrt{\epsilon_k^2+|\Delta_{+-}|^2}$ and $\Delta_{\eta \lambda}$ pair states with the same or different helicities ($\eta, \lambda=\pm$). 

On the other hand if we substitute in Eqs.~\ref{akt} and \ref{bkt}, $\epsilon_k^a \rightarrow \epsilon_k -h$ and $\epsilon_k^b \rightarrow \epsilon_k +h$, we obtain the same result for the energies with the identification~\cite{3dSOC,3dSOC1,alicea,3dSOC2,alicea}, 
\begin{eqnarray}
\label{aliceag}
\Delta_{+-}&=&\frac{h}{\sqrt{h^2+|V_k|^2}}\Delta_{ab} \\ \nonumber
\Delta_{++}&=&\Delta^{\ast}_{--}=\frac{-|V_k|}{ \sqrt{h^2+|V_k|^2}}\frac{k_x+ik_y}{k_{\perp}}\Delta_{ab}.
\end{eqnarray}
It is interesting to notice that the limits $h\rightarrow 0$ and $V$ or $\gamma \rightarrow 0$  ($V=\gamma k_{\perp}$) of the expressions above do not commute. Indeed, for $h \rightarrow 0$ and $V$ finite, we find,
\begin{eqnarray}
\Delta_{+-}&=&0 \\ \nonumber
\Delta_{++}&=&\Delta^{\ast}_{--}=- \frac{k_x+ik_y}{k_{\perp}}\Delta_{ab}
\end{eqnarray}
while, for $V \rightarrow 0$ and $h$ finite, we obtain,
\begin{eqnarray}
\Delta_{+-}&=&\Delta_{ab} \\ \nonumber
\Delta_{++}&=&\Delta^{\ast}_{--}=0.
\end{eqnarray}
This is related to the fact that space inversion and time reversal operations do not necessarily commute~\cite{lee}.
In our approach, that diagonalizes the full Hamiltonian with BCS, SOC and Zeeman terms such ambiguity does not arise.
The relation $ |\Delta_{++}|^2 +|\Delta_{+-}|^2=|\Delta_{ab}|^2$ which follows from Eqs.~\ref{aliceag} implies that the order parameter $\Delta_{ab}$  used here  has contributions from pairing both the same and different  helicity states.

\section{Conclusions}

We have studied the effects of hybridization on superconductivity in a two-band system with inter-band interactions. We focused in the limit of small mismatches between the Fermi wave-vectors of these bands, where the system is always a superfluid at $T=0$. 

Hybridization is a key concept in chemistry and solid state physics. In the latter case it arises from the mixing of different orbitals by the crystalline potential. It can occur locally, at an atomic site, for non-orthogonal wave-functions, as in the case of $s-d$ and $s-f$ mixing. Here the $s$-state is a plane wave containing all the harmonics. Also, it takes place between orbitals in neighboring sites and in this case mixing can involve generic orbitals. Most interesting, as we have shown here,  is  when it occurs in neighboring sites  between orbitals with different parities, as for these with angular momentum \textit{l} and \textit{l+1}, like for $p-d$ and $d-f$ orbitals. In this case the $k$-dependent hybridizations like $V_{pd}(k)$ and $V_{df}(k)$ are not invariant under space inversion symmetry, with the anti-symmetric property, $V_{pd}(-k)=-V_{pd}(k)$ or $V_{df}(-k)=-V_{df}(k)$ even for inversion symmetric lattices. As we have shown this property of the hybridization has dramatic effects on superconductivity where the BCS interaction mixes states with opposite momenta~\cite{act}.  We have shown that anti-symmetric hybridization enhances superconductivity and drives the BCS-BEC crossover even at weak coupling. As mixing among the orbitals can be tuned by doping or external pressure, this turns out to be a controllable mechanism for enhancement of superconductivity.  Besides, this provides an important parameter to explore the quantum phase diagrams of systems where hybridization is anti-symmetric. This includes classes of systems which are of great interest as the transition metal oxides in the case of $V_{dp}$ and heavy fermions for $V_{df}$ hybridization.

We have also shown that the two-band problem with anti-symmetric hybridization is formally equivalent to that of a single band system polarized by an external magnetic field  with a spin-orbit Rashba coupling between the spin up and down bands.
This is a useful analogy as many concepts from one field can be easily brought to the other.

For completeness, we have also studied the effect of symmetric hybridization in two-band superconductivity in both weak and strong coupling regimes. We have shown this acts in detriment of superconductivity and gives rise to quantum phases transition from the superfluid to a normal state.

{\ack  MAC would like to thank Claudine Lacroix for invaluable discussions during a stay in Grenoble financed by the CAPES-COFECB program. We thank Ricardo Sonego Farias and Francisco Dinola-Neto for useful discussions.   HC  acknowledges CBPF for the kind  hospitality. We wish to thank the Brazilian agencies, FAPEAM,
FAPERJ, CNPq and FAPEMIG for financial support}


\begin{thebibliography}{99}
\bibitem{act} The $V_{df}(k)$ hybridization is essential to understand the magnetism of the series of the  actinides (see Refs.~\cite{coqblin,castroneto}). However when studying magnetic properties, the anti-symmetry property of $V_{df}(k)$ does not play a fundamental role (see Refs.~\cite{coqblin,castroneto}).
\end{thebibliography}
\end{document}